\documentstyle[amssymb,prb,aps,epsfig,preprint]{revtex}  
\begin{document}
\title{Itinerant Ferromagnetism and Quantum Criticality in Sc$_3$In}
\author{A. Aguayo$^1$ and D.J. Singh}
\address{Center for Computational Materials Science,
Naval Research Laboratory, Washington, DC 20375 \\
$^1$ also at School of Computational Sciences, George Mason University,
Fairfax, VA 22030}
\date{\today}
\maketitle

\begin{abstract}
The electronic structure and magnetic properties of
hexagonal Sc$_3$In
are calculated within density functional theory. We find that
the Fermi energy lies in a region of flat Sc $d$ derived bands
leading to a peak in the density of states and Stoner ferromagnetism.
The calculated local spin density and generalized gradient approximation
spin magnetizations are both enhanced with respect to experiment,
which is an indication of significant quantum critical fluctuations,
neglected in these approximations. We find, as expected, that the
ferromagnetism is initially enhanced under pressure, meaning that
the critical point cannot be reached with modest pressure. However,
we find that the density of states peak around
the Fermi energy and the calculated density functional
magnetic properties are sensitive to the $c/a$ ratio, so that
the quantum critical point may be reached under uniaxial strain.
\end{abstract}

\newpage

A number of recent discoveries has lead to
resurgence of interest in the properties of clean
metals near quantum critical points.
These include the discovery of a metamagnetic quantum critical
point in Sr$_3$Ru$_2$O$_7$, \cite{grigera}
the appearance of superconductivity at the antiferromagnetic
critical point in CePd$_2$Si$_2$, \cite{mathur}
presumably unconventional superconductivity
on the ferromagnetic side of the critical point in UGe$_2$,
\cite{saxena} the triplet superconductivity
of Sr$_2$RuO$_4$, \cite{maeno}
(the origin of superconductivity is unknown
in this material, but it has been argued that it is close to
ferromagnetic and antiferromagnetic critical points \cite{mazin-singh})
and the recently discovered superconductivity in ZrZn$_2$. \cite{pfleiderer}
These discoveries suggest investigation of other clean metallic compounds with
weak itinerant ferromagnetism that might be driven to the critical
point, {\it e.g.} under moderate pressure.
Common features of these metallic ferromagnets
are: a low saturation moment at 0 K, low
magnetic ordering temperature,
high magnetic susceptibility at 0 K and
negative magnetoresistance.
The essential feature is that the physics are dominated down to quite
low temperature (at the critical point down to 0 K) by quantum
fluctuations associated with the critical point - these are manifest
in the magnetic, transport and thermodynamic properties, {\it e.g.}
as non-Fermi liquid scalings. Thus, in searching for such systems,
we seek materials that are near the critical point with respect to
some experimentally variable parameter (like pressure) and that
are magnetically soft so that the critical region has a reasonable
extent.

Early on, it was reported that Sc$_3$In is such a weak itinerant
ferromagnet that may be near such a critical point. \cite{matthias}
However, in Sc$_3$In under pressure, the magnetic moment is enhanced,
so that the critical point cannot be reached at least with
modest compression. \cite{gardner,grewe,riedi}
Here we report
density functional studies that support the view that Sc$_3$In
is indeed near a quantum critical point, and show that, while
the ferromagnetism is enhanced under pressure, it is
suppressed by uniaxial strain, suggesting non-hydrostatic
pressure as a practical way to reach the quantum critical point.

The calculations were done using the general potential
linearized augmented planewave (LAPW) method, \cite{singh-book}
as implemented
in the WIEN2k code.\cite{wien}
Relativistic effects were included at the scalar relativistic
level. However, we verified that the magnetic moment at the
experimental structure is not sensitive to the inclusion of spin-orbit.
This is presumably because the band structure near the Fermi energy
is dominated by Sc $d$ bands (see below).
For the generalized gradient approximation (GGA) calculations,
we used the exchange-correlation functional of Perdew, Burke and Ernzerhof.
\cite{pbe96}
We chose the muffin-tin spheres $R_{MT} = 2.6$ a.u. and
a basis set determined by a planewave cutoff of $R_{MT}K_{max}=8.0$, which
gives good convergence. The Brillouin zone samplings
were done using the special {\bf k} points method with
528 points in the irreducible wedge of the hexagonal zone.
Convergence was checked by varying the {\bf k} point density in
the zone averages. Convergence was further checked by repeating some
calculations
with a different LAPW code \cite{wei,singh-lo} using different sphere radii,
local orbitals to relax linearization \cite{singh-lo}
and a higher planewave cutoff, $R_{MT}K_{max}=9.0$.
This yielded essentially the same results.

Hexagonal Sc$_3$In occurs in the
Ni$_3$Sn-type structure
(spacegroup $P6_3/mmc$).\cite{palenzona}
The experimental lattice parameters are $a=6.43$ \AA~ and $c=5.17$ \AA.
\cite{lat-parm}
In this structure, the space group symmetry allows a free coordinate $x$,
which has not been fixed experimentally and
which gives the in-plane position of the Sc.
However, it is expected based
on solid state chemical reasoning (particularly the Sc coordination)
that Sc should sit on the ideal site, $x$=5/6.
We performed local density approximation and GGA
calculations of the energy as a function of $x$ to check this,
and found that, to within the precision of our calculations,
$x$ indeed takes the ideal value of 5/6 in Sc$_3$In
(the energy minimization yielded $x$=0.832); the calculated
magnetic moment was found to be only weakly sensitive to changes
in $x$ of this order).

\begin{figure}[tbp]
\centerline{\epsfig{file=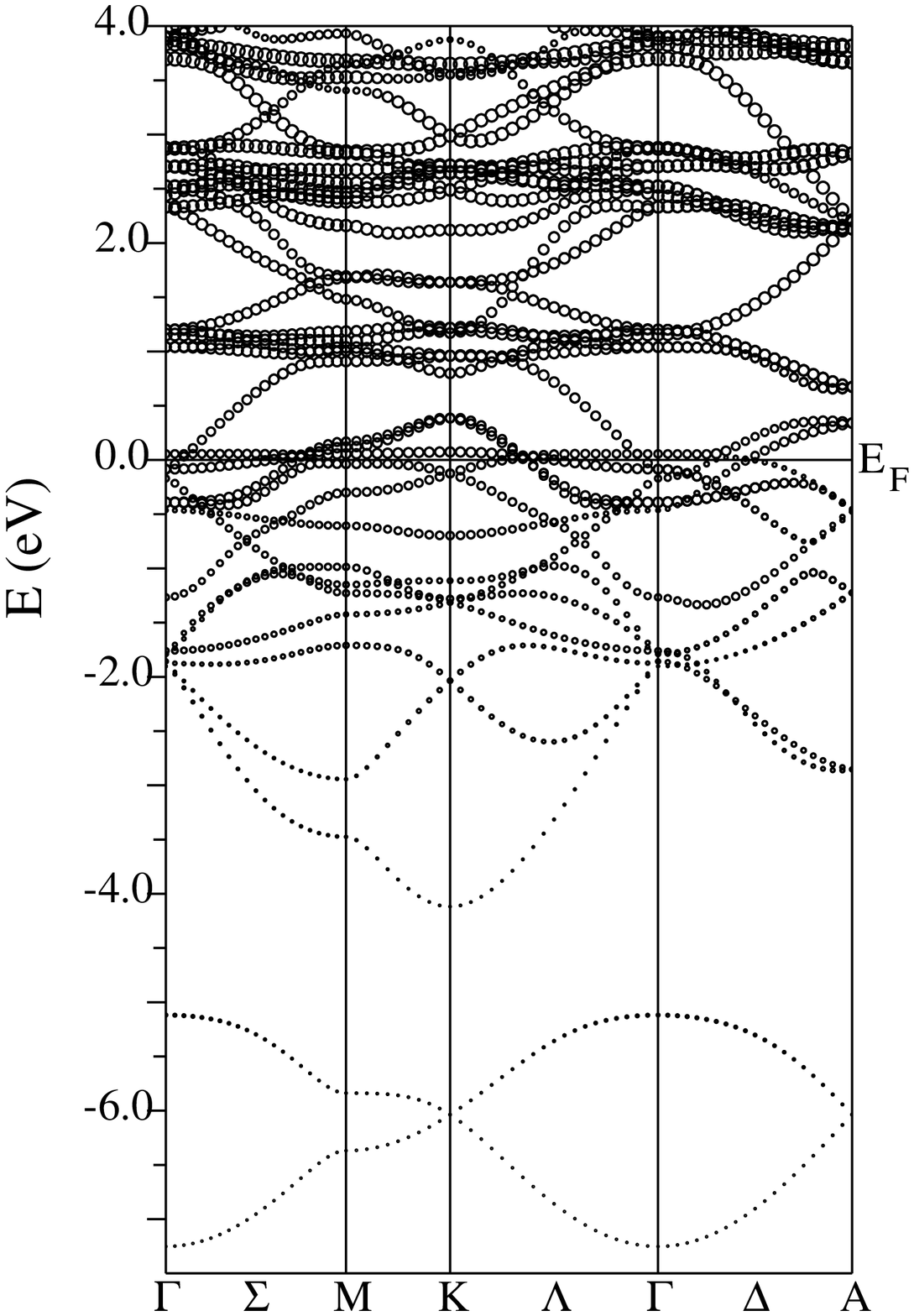,width=0.80\linewidth,clip=}}
\vspace{-0.3cm}
\centerline{\epsfig{file=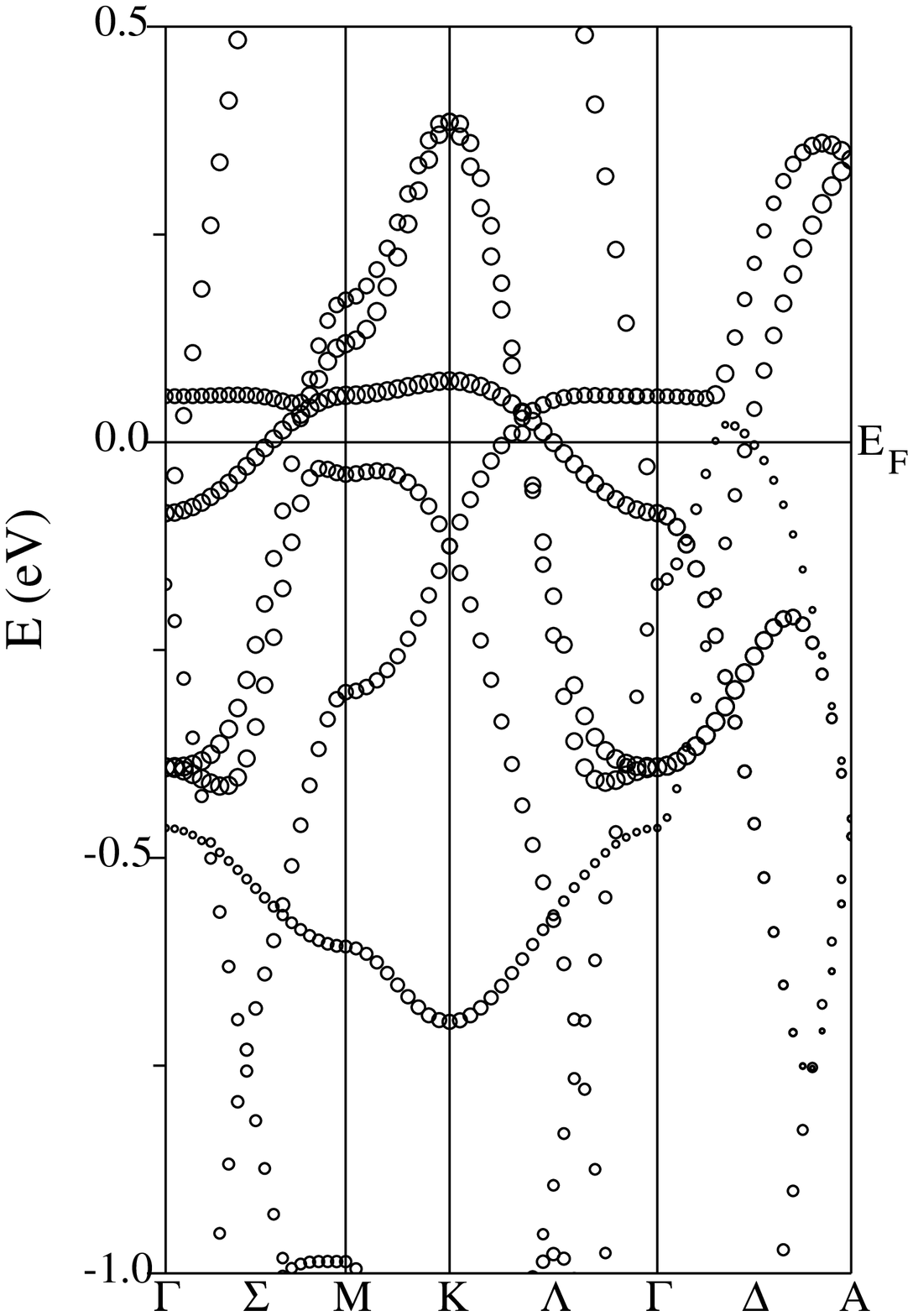,width=0.80\linewidth,clip=}}
\vspace{-0.3cm}
\caption{Valence band structure of Sc$_3$In (top) and a blow-up
around $E_F$ (bottom).
The radius of the plotting
circles is proportional to Sc $d$ projection of the band.}
\label{spa}
\end{figure}

\begin{figure}[tbp]
\centerline{\epsfig{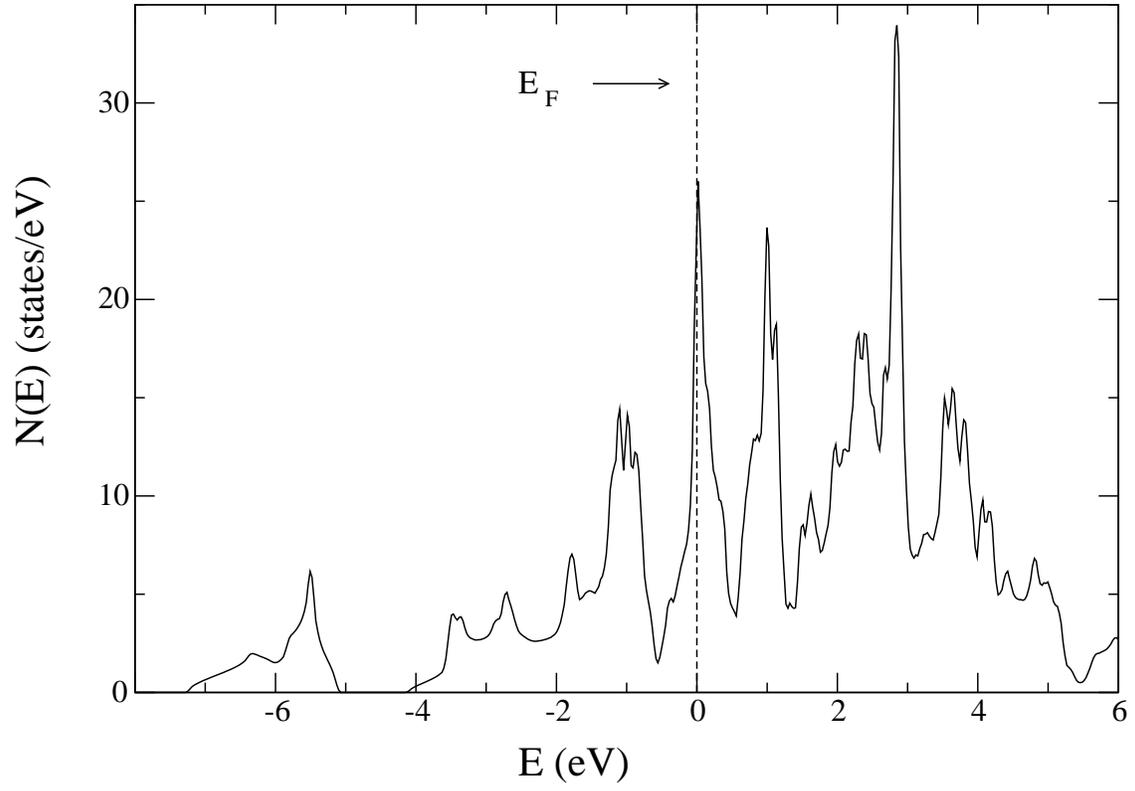}}
\vspace{0.2cm}
\vspace{0.2cm}
\caption{Calculated electronic DOS on a per
unit cell basis of non-spin-polarized Sc$_3$In
obtained within the GGA. The zero is at $E_F$.}
\label{dos-fig}
\end{figure}

We show the paramagnetic electronic band structure
calculated at
the experimental crystal structure
with the ideal value of $x$ in Fig. \ref{spa},
along with a blow-up of the
region around $E_F$. The corresponding electronic density of states
is given in Fig. \ref{dos-fig}.
In the band structure, the Sc $d$ character is indicated by the relative size
of the plotting symbols.
The two lowest valence bands lying between approximately
-7 eV and -5 eV with respect to
the Fermi energy, $E_F$, are In $s$ derived, while the remaining valence bands
beginning at -4 eV are formally described as Sc-In hybrids. However,
from approximately -1 eV to several eV above $E_F$, these are heavily
dominated by Sc $d$ states. The band structure near $E_F$ shows two
very flat Sc $d$ derived bands as well as three more dispersive bands
of mixed Sc $d$ - In $p$ character that pass through them.
The flat
bands lead to a prominent peak in the density of states (DOS) around
$E_F$. The GGA DOS at $E_F$ is $N(E_F)=26.5$ eV$^{-1}$ per unit cell
(two formula units).

This peak underlies the itinerant ferromagnetism of Sc$_3$In,
which occurs via the itinerant Stoner mechanism.
We indeed find a ferromagnetic ground state in self-consistent calculations.
The calculated GGA spin moment with the experimental structure is
3.0 $\mu_B$ per unit cell ({\it i.e.} 0.50 $\mu_B$/Sc).
Again with the experimental crystal structure, local
spin density approximation (LSDA) calculations give a smaller
moment of 2.1 $\mu_B$ also on a per unit cell basis
(0.35 $\mu_B$/Sc).
The magnetic energies, defined as the energy difference
between constrained non-spin-polarized and ferromagnetic states,
are very small -- 1.3 mRy/Sc in the GGA
and only 0.2 mRy/Sc in the LSDA.
It is not unusual for GGA calculations to yield a somewhat greater
tendency towards magnetism than the LSDA, \cite{ashkenazi} but
the magnitude of the difference here is large suggesting
longitudinal magnetic softness.
We note that both the LSDA and GGA spin magnetizations are
considerably larger than experiment.
Mattthias {\it et al.} reported a magnetic moment of 0.051 $\mu_B$/Sc,
which is
close to the value of 0.066 $\mu_B$/Sc found by Gardner {\it et al.}
(Ref. \onlinecite{gardner}).
This level of disagreement may be taken as indicating
the importance of quantum critical fluctuations in Sc$_3$In.

Density functional theory is in principle an exact ground state
theory. It should, therefore, correctly describe the spin density
of magnetic systems. However, common approximations to the exact density
functional theory, such
as the local spin density approximation (LSDA) and generalized gradient
approximations (GGA), neglect Hubbard correlations beyond the
mean field level, with the well known result that the magnetic
tendency of strongly Hubbard correlated systems is often underestimated.
Another type of correlations that are missed in these approximations
are quantum spin fluctuations. This is because the LSDA and GGA are
parameterized based on electron gasses with densities typical for atoms and
solids. However, the uniform electron gas is very far from magnetism in
this density range. The result for non-collinear magnetism in molecules is
well known - a so called ``spin-contamination'' in which the LSDA predicts
classical magnetism, with {\it e.g.} non-quantized magnetic moments
and singlets described with non-zero, static, but canceling, spin
densities on various sites. In solids near quantum critical points,
the result is an overestimate of the magnetic moments and tendency
towards magnetism ({\it i.e.} misplacement of the
position of the critical point) due to neglect
of the quantum critical fluctuations. Since it is very uncommon for
the LSDA to significantly overestimate the tendency towards magnetism
away from such critical points, we take the overestimate found here as
evidence for the importance of quantum critical fluctuations in Sc$_3$In.
A similar overestimate was reported recently for ZrZn$_2$, \cite{singh-mazin2}
which shows a critical point under pressure and coexistence of ferromagnetism
and superconductivity. \cite{pfleiderer} Similarly, in Sr$_3$Ru$_2$O$_7$,
LSDA calculations predict a weakly ferromagnetic ground state, \cite{singh-327}
while experimentally the material is a paramagnet very close to ferromagnetism.
\cite{grigera}
Unfortunately, in Sc$_3$In, the ordinary method of reaching the critical
point by applying hydrostatic pressure does not work, since,
as mentioned, it is known
that the magnetism of this compound becomes more robust under modest pressures.
\cite{gardner,grewe,riedi}

\begin{figure}[tbp]
\centerline{\epsfig{file=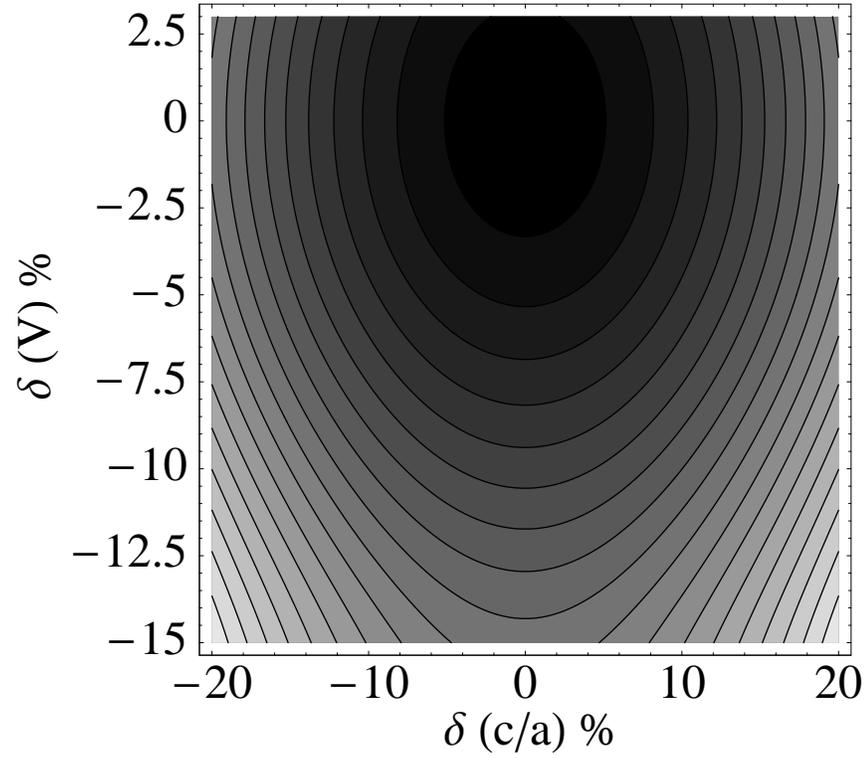,width=0.95\linewidth,clip=}}
\vspace{-2.5cm}
\caption{Total energy surface as function of $c/a$ and volume.
$\delta(V)$ and $\delta(c/a)$ are the percentage deviations of
$V$ and $c/a$ from their experimental values.
The contours are spaced by 10 mRy per unit cell.}
\label{eos}
\end{figure}

In order to investigate this further and to see if there is another
way of reaching the critical point, we performed GGA calculations of the
total energy, magnetic moment and electronic structure as
a function of the structural parameters $c$ and $a$. These calculations
were performed on a search grid with volume changes from -15\% to
+3\% and $c/a$ ratios from -40\% to +40\%, both with respect
to the experimental structure.
The internal parameter $x$ was kept fixed at its ideal value during these
calculations.
A contour representation of the energy surface is given in
Fig. \ref{eos}.
The calculated GGA structural parameters
are in very close agreement with experiment.
These were obtained by
fitting the total energy surface $E(c/a, V)$ to a polynomial.
The resulting equilibrium volume is $V= 187$ \AA$^3$ per unit cell with
the calculated $c/a=0.809$.
The experimental values are
$V=185$ \AA$^3$ and $c/a = 0.804$.\cite{grewe}
The calculated GGA constrained bulk modulus is 64 GPa (this is not
strictly the same as the experimental bulk modulus as it is from a
fit with a fixed $c/a$ ratio, while experimentally $c/a$ changes under
hydrostatic pressure).

\begin{figure}[tbp]
\centerline{\epsfig{file=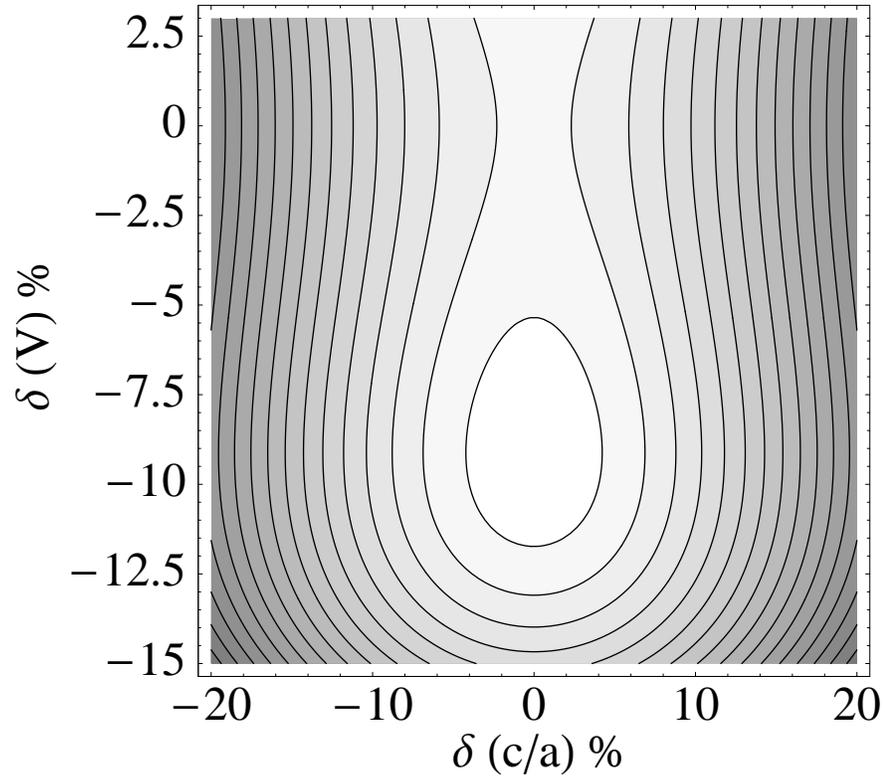,width=0.95\linewidth,clip=}}
\vspace{-2.5cm}
\caption{Variation of the magnetic moment with $c/a$ and volume
as in Fig. \ref{eos}.
The contours are spaced by 0.10 $\mu_B$ per unit cell.}
\label{magsur}
\end{figure}

The variation of the calculated GGA spin
moment as a function of $V$ and $c/a$ is shown in Fig. \ref{magsur}.
The magnetic moment increases weakly under modest compression
in agreement with the experimental trend, \cite{gardner} and then
decreases, but only weakly, at high compressions.
Volume compressions above 16\% are needed before the GGA spin magnetization
falls below its zero pressure value.
The variation with $c/a$ is stronger and
more interesting. In particular, we find that the GGA
magnetization has a maximum very close to the experimental $c/a$,
which is also the $c/a$ ratio for which $E_F$ is in the
center of the two heavy bands
giving rise to the DOS peak that underlies magnetism. As $c/a$ is
changed these bands shift with respect to $E_F$ and so as expected
from extended Stoner theory the spin moment and magnetic energy is
reduced. \cite{krasko}

Considering the importance of quantum critical fluctuations, which
strongly reduce the GGA magnetization, we cannot predict the position
of the critical point. However, it is clear that Sc$_3$In is close
to the critical point and that the magnetism can be suppressed with
variations in the $c/a$ ratio. Accordingly, we expect that the critical point
can be reached under modest uniaxial strain. Experimental investigation
of the low temperature magnetic, thermodynamic and
transport properties of high quality single crystals of Sc$_3$In under
uniaxial strain should be quite interesting.

We are grateful for helpful conversations with A.J. Millis, D. Mandrus
I.I. Mazin and B.C. Sales.
Work at the Naval Research Laboratory is
supported by the Office of the Naval Research.

\end{document}